\documentclass[a4paper,12pt]{article}
\usepackage{graphics}
\usepackage{amsfonts}

\newlength{\extraspace}
\newlength{\extraspaces}
\catcode`\@=11
 
\def\numberbysection{\@addtoreset{equation}{section}
\def\theequation{\arabic{section}.\arabic{equation}}}

\begin{document}
%
\thispagestyle{empty}
\begin{center}
\begin{flushright}
TIT/HEP--431 \\
{\tt hep-th/9910095} \\
October, 1999 \\
\end{flushright}
\vspace{3mm}
\begin{center}
{\Large
{\bf An Exact Solution of BPS Domain Wall Junction
}} 
\\[18mm]

{\sc Hodaka~Oda},\footnote{
\tt e-mail: hoda@th.phys.titech.ac.jp} \hspace{2.0mm}
{\sc Kenji~Ito},\footnote{
\tt e-mail: kito@th.phys.titech.ac.jp} \hspace{2.0mm}
{\sc Masashi~Naganuma},\footnote{
\tt e-mail: naganuma@th.phys.titech.ac.jp} \hspace{2.0mm}
and \hspace{2.0mm}
{\sc Norisuke~Sakai}\footnote{
\tt e-mail: nsakai@th.phys.titech.ac.jp} \\[3mm]
{\it Department of Physics, 
Tokyo Institute of Technology \\[2mm]
Oh-okayama, Meguro, Tokyo 152-8551, Japan} \\[4mm]

\end{center}
\vspace{18mm}
{\bf Abstract}\\[5mm]
{\parbox{13cm}{\hspace{5mm}
An exact solution of domain wall junction is obtained in a 
four-dimensional ${\cal N}=1$ supersymmetric $U(1)\times U(1)'$ gauge 
theory 
with three pairs of chiral superfields which is motivated by the 
${\cal N}=2$ $SU(2)$ gauge theory with one flavor perturbed by 
an adjoint scalar mass. 
The solution allows us to evaluate various quantities including 
a new central charge $Y_k$ associated with the junction 
besides $Z_k$ which appears already in domain walls. 
We 
find that the new central charge $Y_k$ 
gives a negative contribution to the mass of the domain wall junction 
whereas the central charge $Z_k$ gives a dominant positive contribution. 
One has to be cautious to identify 
the central charge $Y_k$ alone as the mass of the junction. 
}}
\end{center}
\vfill
\newpage
\vfill
\newpage
\setcounter{equation}{0}
\setcounter{footnote}{0}

{\large { \bf Introduction}}

\vspace{5mm}

In recent years, there has been an intensive study of domain walls which 
appear in many areas of physics.
These domain walls interpolate between degenerate discrete minima of 
a potential and spread over two spatial dimensions. 
This situation arises naturally in four-dimensional ${\cal N}=1$ 
supersymmetric 
field theories \cite{DvaliShifman}--\cite{KSY} 
in addition to condensed matter physics. 
In supersymmetric unified models, 
domain walls can be formed during thermal evolution of our universe 
and often provide significant and interesting 
constraints on model building. 
On the other hand, it has been found that domain walls in supersymmetric 
theories can saturate the Bogomol'nyi bound \cite{BPS}. 
Such a domain wall preserves half of the original supersymmetry and is 
called $1/2$ BPS state \cite{WittenOlive}. 
It has also been noted that these BPS states possess a topological charge 
which becomes a central charge $Z$ of the supersymmetry algebra 
\cite{AbrahamTownsend} \cite{DvaliShifman}. 

Recently another interesting possibility for a BPS state has attracted 
much attention \cite{GibbonsTownsend}--\cite{GorskyShifman}. 
Domain walls occur in interpolating two discrete degenerate vacua 
in separate region of space. 
If three or more different discrete vacua occur in separate 
region of space, segments of domain walls separate 
each pair of the neighboring vacua. 
If the two spatial dimensions of all of these domain walls have 
one dimension in common, 
these domain walls meet at a one-dimensional junction. 
The solitonic configuration for the junction can 
preserve a quarter of supersymmetry. 
It has also been found that a new topological charge $Y$ 
can appear for such a $1/4$ BPS state \cite{GibbonsTownsend} \cite{CHT} 
\cite{GorskyShifman}. 

There have been general considerations of junctions \cite{GibbonsTownsend} 
\cite{CHT}  as well as 
more concrete numerical results \cite{Saffin}. 
In spite of these efforts, 
no exact or explicit solution has been obtained so far 
for the BPS junctions. 
In order to make progress in understanding these solitonic objects, 
it is quite useful to have exact solutions which allows us to 
investigate closely the behavior of these solitons and to 
evaluate explicitly the central charges $Y$ besides $Z$. 
In this respect, an exact solution offers informations complementary to 
general considerations and numerical studies. 

The purpose of our paper is to present an exact solution of domain wall 
junction with three distinct vacua in a field theory model and 
explicitly work out various properties 
of the soliton including the new central charge $Y$ as well as the central 
charge $Z$. 
We believe that this is the first exact analytic solution of 
the BPS domain wall junction. 
The model is a simplified toy model simulating 
the ${\cal N}=2$ supersymmetric $SU(2)$ Yang-Mills theory with one flavor 
which is explicitly broken to ${\cal N}=1$ by giving a mass term to the 
adjoint chiral superfield. 
The 
central charge $Z$ is a two-dimensional complex 
vector which is determined by differences of superpotential 
at three distinct vacua. 
We give a formula which explicitly expresses the energy of the domain walls 
and junctions in terms of the central charges $Z$ and $Y$. 
We find in our model that the central charge $Y$ has a simple geometrical 
meaning of the $-2$ times the triangular area in field space which is 
enclosed by three domain walls connecting three distinct vacua at infinity. 
We also find that the 
main contribution to the mass of the domain wall junction configuration 
comes from the central charge $Z$ and the negative $ Y$ 
is merely an additional small negative contribution. 
Our result gives a warning to a naive identification of 
the central charge $Y$ alone to be the mass of the junction. 

\vspace{5mm}
{ \bf Junctions and Central Charge}

\vspace{5mm}

Using the convention of ref.\cite{WessBagger},  we denote 
the left-handed and right-handed supercharges of 
the ${\cal N}=1$ supersymmetric four-dimensional field theory 
as $Q_\alpha$, $\bar{Q}_{\dot{\alpha}}$ . 
If the translational invariance is broken as is the case for domain 
walls and/or junctions, the superalgebra in general receives 
contributions from central charges \cite{DvaliShifman}, 
\cite{AbrahamTownsend}--\cite{CHT}, 
\cite{GorskyShifman}. 
The anti-commutator between two left-handed supercharges has central 
charges 
$Z_k$,  $k=1,2,3$ 
\begin{equation}
\{ Q_\alpha, Q_\beta  \} =
2 i (\sigma^k \bar{\sigma}^0 )_\alpha{}^{\gamma} 
\epsilon_{\gamma \beta} Z_k. 
\label{qqantcom}
\end{equation}
The anti-commutator between left- and right-handed supercharges 
receives a contribution from central charges $Y_k, \ k=1,2,3$  
\begin{equation}
\{ Q_\alpha, \bar{Q}_{\dot{\alpha}} \} =
2 (\sigma^{\mu}_{\alpha \dot{\alpha}} P_{\mu} 
+ \sigma^k_{\alpha \dot{\alpha}} Y_k ),
\label{qqantcomy}
\end{equation}
where $P_{\mu}, \ {\mu}=0, \cdots, 3$ are the energy-momentum 
four-vector of 
the system. 
Hermiticity of supercharges dictates that the central charges $Z_k$ 
are complex, and that $Y_k$ are real: $(Y_k)^*=Y_k$. 

These central charges come from the total divergence and are 
non-vanishing when there are nontrivial differences in asymptotic 
behavior in different region of spatial infinity as is the case of domain 
walls and junctions. 
Therefore these charges are topological in the sense that they are 
determined completely by the boundary conditions at infinity. 
{}For instance, we can compute the anticommutators (\ref{qqantcom}), 
(\ref{qqantcomy}) in the general Wess-Zumino models with 
arbitrary number of chiral superfields $\Phi^i$ and arbitrary superpotential 
$W$. The contributions to the central charges from bosonic components of 
chiral superfields are given by 
\begin{equation}
Z_k = 2 \int d^3 x \, \partial_k W^*(A^*),
\label{centralchargeZ}
\end{equation}
\begin{equation}
Y_k =i \epsilon^{knm}
\int d^3 x \, K_{i j^*}
\partial_n (A^{*j} \partial_m A^i), \qquad  \epsilon^{123}=1, 
\label{centralchargeY}
\end{equation}
where the scalar component of the $i$-th chiral superfield $\Phi^i$ is 
denoted as $A^i$ and the K\"ahler metric 
$ K_{i j^*}=\partial^2 K(A^*, A) /\partial A^i\partial A^{*j}$ 
is obtained from the K\"ahler potential $K$. 
We see that the central charge 
$ Z_k $ is completely determined by 
the difference of values of the superpotential $W$ at spatial infinities 
where different discrete vacua are chosen for different directions. 
Since single domain wall has a field configuration which is nontrivial  
only in one dimension, one can see from eq.(\ref{centralchargeY}) that 
the central charge $Y_k$ vanishes whereas the central charge $Z_k$ 
is non-vanishing. 
The central charge $Y_k$ can be non-vanishing, 
if the field configuration at infinity is nontrivial in two-dimensions. 
This situation occurs 
when three or more different vacua occur at infinity as is 
the case for the domain wall junctions. 

To examine the lower bound for the energy due to the hermiticity of the 
supercharges, we consider a hermitian linear combination of operators 
$Q$ and $\bar{Q}$ with an arbitrary complex two-vector $\beta^\alpha$ 
and its complex conjugate $\bar{\beta}^{\dot{\alpha}} = (\beta^\alpha)^*$ 
as coefficients 
\begin{equation}
K= \beta^\alpha Q_\alpha + 
\bar{\beta}^{\dot{\alpha}} \bar{Q}_{\dot{\alpha}}. 
\end{equation}
We treat $\beta^\alpha$ as c-numbers rather than
the Grassmann numbers. 
Since $K$ is hermitian, 
the expectation value of the square of $K$ over any state 
is non-negative definite 
\begin{eqnarray}
\langle S | K^2 | S \rangle 
&\!\!\! \equiv &\!\!\! 
(\beta^1 , \beta^2, \bar{\beta}^{\dot{1}}, \bar{\beta}^{\dot{2}})
\hat{K}^2 
\left(
 \begin{array}{c}
 \beta^1 \\
 \beta^2 \\
 \bar{\beta}^{\dot{1}} \\
 \bar{\beta}^{\dot{2}}
 \end{array}
\right) \ge 
0. 
\label{ineq}
\end{eqnarray}
The equality holds if and only if the linear combination of supercharges 
$K$ is preserved by the state $ |S \rangle$. 
Since we are interested in field configurations at rest, we obtain 
$P^k=0, (k=1,2,3)$ 
and the matrix $\hat{K}^2$ in terms of the central charges $Z_k$, $Y_k$ 
and the hamiltonian $H$ explicitly 
\begin{equation}
\hat{K}^2 =
\left(
 \begin{array}{cccc}
  \langle -Z_2 - i Z_1 \rangle  & \langle i Z_3 \rangle &
    \langle H + Y_3 \rangle   & \langle Y_1 - i Y_2  \rangle \\
  \langle i Z_3 \rangle & \langle -Z_2+i Z_1 \rangle   &
    \langle Y_1+i Y_2 \rangle & \langle H - Y_3      \rangle \\
  \langle H + Y_3 \rangle   & \langle Y_1 + i Y_2  \rangle &
    \langle -Z_2^* + i Z_1^* \rangle  & \langle -i Z_3^* \rangle \\
  \langle Y_1-i Y_2 \rangle & \langle H - Y_3 \rangle   &
  \langle -i Z_3^* \rangle  & \langle -Z_2^*-i Z_1^* \rangle   
 \end{array}
\right).
\end{equation}

{}For simplicity, let us assume that field configuration is two-dimensional, 
for instance, depends on $x_1$, $x_2$ only. 
Then we obtain 
$\langle Z_3 \rangle=\langle Y_1 \rangle=\langle Y_2 \rangle = 0$. 
The inequality (\ref{ineq}) implies in this case 
that for any $\beta$ and any state 
\begin{eqnarray}
\langle H \rangle
&\!\!\! \ge &\!\!\!
\frac{-1}{|\beta^1|^2+|\beta^2|^2} \Biggl\{
(|\beta^1|^2-|\beta^2|^2) \langle Y_3 \rangle
+ {\rm Re} \left[(\beta^1)^2 \langle - Z_2 -i Z_1 \rangle \right]
\nonumber \\
&\!\!\!  &\!\!\! {}
+ {\rm Re} \left[(\beta^2)^2 \langle - Z_2 +i Z_1 \rangle \right]
\Biggr\}.
\end{eqnarray}
The minimum energy is achieved at the larger one of vanishing eigenvalues 
of the matrix $\hat{K}^2$ 
\begin{equation}
{\rm det}(\hat{K}^2)=(\langle H+Y_3 \rangle^2 
- |\langle -i Z_1-Z_2 \rangle|^2 )
(\langle H-Y_3 \rangle^2 
- |\langle i Z_1-Z_2 \rangle|^2 )
=0.
\label{vanishingdet}
\end{equation}
Thus the BPS bound becomes 
$
\langle H \rangle \ge 
{\rm max}\{ 
H_{\rm I} , H_{\rm II}
\}
$ 
where $H_{\rm I}$ and $H_{\rm II}$ are two solutions of 
eq.(\ref{vanishingdet}) 
\begin{equation}
 H_{\rm I} \equiv
|\langle -i Z_1-Z_2 \rangle|
-\langle Y_3 \rangle, 
\quad 
 H_{\rm II} \equiv
|\langle i Z_1-Z_2 \rangle|
+\langle Y_3 \rangle. 
\label{BPSenerges}
\end{equation}
The corresponding eigenvectors are given by 
$
\bar{\beta}^{\dot{1}}=\beta^1
\langle i Z_1+Z_2 \rangle/
{|\langle i Z_1+Z_2 \rangle|}, \ \ 
\beta^2=\bar{\beta}^{\dot{2}}=0
$ for $\langle H \rangle = H_{\rm I}$ and  
$
\beta^1=\bar{\beta}^{\dot{1}}=0, \ \
\bar{\beta}^{\dot{2}}=\beta^2
\langle -i Z_1+Z_2 \rangle/
{|\langle -i Z_1+Z_2 \rangle|}
$ for 
$
\langle H \rangle = H_{\rm II} 
$
.

If $H_{\rm I} > H_{\rm II}$, 
then supersymmetry can only be preserved at $\langle H \rangle=H_{\rm I}$ 
and 
the only one combination of supercharges is  conserved 
\begin{equation}
\left(
Q_1 + 
\frac{\langle i Z_1+Z_2 \rangle}
{|\langle i Z_1+Z_2 \rangle|}
\bar{Q}_{\dot{1}}
\right)
| {\rm BPS} \rangle =0.
\label{conseved_charge_1st}
\end{equation}
If $H_{\rm II}>H_{\rm I}$, 
then supersymmetry can only be preserved at 
$\langle H \rangle = H_{\rm II}$ and
the only one combination of supercharges is  conserved 
\begin{equation}
\left(
Q_2 + 
\frac{\langle -i Z_1+Z_2 \rangle}
{|\langle -i Z_1+Z_2 \rangle|}
\bar{Q}_{\dot{2}}
\right)
| {\rm BPS} \rangle =0.
\label{conseved_charge_2st}
\end{equation}
These cases correspond to the $1/4$ BPS state. 
If two eigenvalues are degenerate $H_{\rm I}=H_{\rm II}$, 
we can have $1/2$ BPS state at $H=H_{\rm I}=H_{\rm II}$ 
where both two combinations 
of supercharges (\ref{conseved_charge_1st}) and (\ref{conseved_charge_2st}) 
are conserved. 

The condition of supercharge conservation (\ref{conseved_charge_1st}) 
for $H=H_{\rm I}$ applied to chiral superfield 
$\Phi^i=(A^i, \psi^i, F^i)$ gives 
after eliminating the auxiliary field $F^i$ 
\begin{equation}
2 i K_{i j^*} 
\frac{\langle i Z_1+Z_2 \rangle}
{|\langle i Z_1+Z_2 \rangle|}
{\cal D}_{\bar{z}} A^i
= - \frac{\partial W^*}{ \partial A^{*j}}, 
\label{Be1}
\end{equation}
where complex coordinates $z=x_1+i x_2, \bar{z}=x_1-i x_2$, gauge covariant
derivatives ${\cal D}_\mu $, ${\cal D}_{\bar{z}}=\frac{1}{2}({\cal D}_1+i {\cal D}_2)$ and ${\cal D}_z=\frac{1}{2}({\cal D}_1-i {\cal D}_2)$ are introduced.
The same BPS condition (\ref{conseved_charge_1st}) applied to $U(1)$ vector 
superfield in the Wess-Zumino gauge $V=(v_\mu, \lambda, D)$ gives 
after eliminating the auxiliary field $D$  
\begin{equation}
v_{12}=-D={1 \over 2} \sum_j A^{*j} e_j A^j, \quad v_{03}=0, \quad 
v_{01}=v_{31}, \quad v_{23}=-v_{02}, 
\label{Be1vector}
\end{equation}
where $v_{\mu \nu}\equiv 
\partial_{\mu} v_{\nu} - \partial_{\nu} v_{\mu}$ and $e_j$ is the charge 
of the field $A^j$. 
A similar condition holds in the case of non-Abelian gauge group. 

Similarly the condition of supercharge conservation 
(\ref{conseved_charge_2st}) 
for $H=H_{\rm II}$ applied to chiral superfield  
gives 
after eliminating the auxiliary field 
\begin{equation}
2 i K_{i j^*} 
\frac{\langle i Z_1-Z_2 \rangle}%
{|\langle i Z_1-Z_2 \rangle|}
{\cal D}_z A^i
= - \frac{\partial W^*}{ \partial A^{*j}}. 
\label{Be2}
\end{equation}
The BPS condition (\ref{conseved_charge_2st}) applied to $U(1)$ vector 
superfield in the Wess-Zumino gauge 
gives
\begin{equation}
v_{12}=D=-{1 \over 2} \sum_j A^{*j} e_j A^j, \quad v_{03}=0, \quad 
v_{01}=-v_{31}, \quad v_{23}=v_{02}. 
\label{Be2vector}
\end{equation}
These BPS conditions (\ref{Be1}) and (\ref{Be1vector}) for $H=H_{\rm I}$ 
and (\ref{Be2}) and (\ref{Be2vector}) for $H=H_{\rm II}$ 
 ensure that the configuration is BPS saturated. 

Since $\langle Z_k \rangle$ and 
$\langle Y_k \rangle$ are given by total divergence as shown in 
eqs.(\ref{centralchargeZ}) and (\ref{centralchargeY}), 
they are fixed by boundary condition at 
spatial infinity. 
Therefore the boundary condition determines which of the supercharges 
can be preserved (\ref{conseved_charge_1st}) 
and/or (\ref{conseved_charge_2st}).

Since the BPS states are the minimum energy solution 
for a given boundary condition at infinity, 
they are stable against any fluctuations preserving the boundary condition. 
The domain wall has the minimum energy and 
is stable as long as two different vacua 
occupy the order $R$ region of boundary of large radius $R$. 
The domain wall junction has also the minimum energy and is stable provided 
the three (or more) vacua remain in regions of order $R$
. 

\vspace{5mm}

{\large {\bf The model }}

\vspace{5mm}

There are many field theory models which have
BPS domain wall or junction solutions.
{}First example is the Wess-Zumino model of single chiral scalar field 
$\Phi$ with a polynomial superpotential 
$W= \Lambda^2 \Phi - \frac{1}{n+1} \Lambda^{2-n} \Phi^{n+1}$,
where $n$ is an integer $\ge 2$ and $\Lambda$ is a parameter with the 
dimension of mass. 
This model has $n$ discrete supersymmetric vacua with vanishing vacuum 
energy. 
Therefore one can have domain wall solutions for $n \ge 2$ 
 \cite{DvaliShifman}, 
and the junction solutions for $n \ge 3$ \cite{AbrahamTownsend} -- 
\cite{CHT}, 
interpolating among those vacua. 
Numerical studies have been performed for domain walls and junctions 
in these models \cite{Saffin}. 
However, no explicit analytic solution has been found even 
for domain walls, apart from the simplest case of $n=2$ where a kink 
solution has been known for sometime. 
No explicit solution has been found for more difficult problem of junctions. 

Another example is the ${\cal N}=1$ supersymmetric QCD with $N_f$ flavor 
of quarks 
in the fundamental representation. 
{}For the case of $SU( N_c )$ gauge group, 
it has $N_c-N_f$ discrete supersymmetric vacua \cite{IntriligatorSeiberg}, 
and can have domain wall solutions \cite{DvaliShifman}--\cite{SmilgaVeselov}. 
This model can also be obtained from the 
${\cal N}=2$ supersymmetric QCD by perturbing with a mass term for the 
adjoint chiral superfield. 
It reduces to the ${\cal N}=1$ supersymmetric gauge theory in the 
infinite mass limit, 
whereas it ends up at the singular points of moduli space 
of the ${\cal N}=2$ supersymmetric gauge theory in the limit of vanishing 
adjoint mass \cite{SeibergWitten}
. 
The moduli space of the ${\cal N}=2$ $SU(2)$ supersymmetric Yang-Mills 
theory has two singularities where monopole or dyon becomes massless 
respectively  \cite{SeibergWitten}. 
In order to discuss the model in a simpler setting, Kaplunovsky 
{\it et.~al.~} have proposed a toy model 
which can be treated as a local field theory \cite{KSY}. 
They introduced two pairs of chiral superfields 
 ${\cal M}, \tilde{\cal M}$ and ${\cal D}, \tilde{\cal D}$ 
simulating the monopole, anti-monopole and the dyon, 
anti-dyon of the Seiberg-Witten theory respectively.
Instead of the modulus $u$ of the Seiberg-Witten theory, 
they introduced a linearized analogue $T$ as a neutral chiral superfield. 
The gauge group was chosen as $U(1)\times U(1)'$ simulating electric 
and magnetic gauge group and the quantum number of these chiral 
superfields 
are given by 
\begin{equation}
\begin{array}{cccccc}
      & {\cal M} & \tilde{\cal M} & {\cal D} & \tilde{\cal D} & T \\
U(1)  &  0       &  0             &   1      &  -1            & 0 \\
U(1)' &  1       &  -1            &   1      &  -1            & 0 
\end{array}
\end{equation}
To mimic a massless monopole at $T=\Lambda$ and a massless dyon at 
$T=-\Lambda$, they consider a superpotential 
\begin{equation}
W=(T-\Lambda) {\cal M} \tilde{{\cal M}}
+ (T+\Lambda) {\cal D} \tilde{{\cal D}}
- h^2 T,
\end{equation}
where the coupling parameter $h^2$ replaces the effect of the mass 
for the adjoint chiral superfield. 
Their model has two discrete ${\cal N}=1$ supersymmetric vacua 
\begin{eqnarray}
T
&\!\!\!=&\!\!\! + \Lambda, \quad 
{\cal M}\tilde{\cal M}=h^2, \quad
\left|{\cal M}\right|=
|\tilde{\cal M}
|, \quad
{\cal D}=\tilde{\cal D}=0, \nonumber \\
T
&\!\!\!=&\!\!\! - \Lambda, \quad 
{\cal D}\tilde{\cal D}=h^2, \quad
\left|{\cal D}\right|=
|\tilde{\cal D}
|, \quad
{\cal M}=\tilde{\cal M}=0.
\label{KapVac}
\end{eqnarray}
{}For simplicity, they assumed that the K\"ahler metric of the model
is flat and discussed the domain wall solution 
interpolating between the two vacua. 
{}For the special case of 
$h^2=2 \Lambda^2$, they obtained an analytic solution of the 
domain wall which asymptotes to the vacuum at $T=+\Lambda$ 
for $x \to - \infty$ 
and to the other vacuum at $T=-\Lambda$ for $x \to + \infty$ 
 of (\ref{KapVac}):
\begin{eqnarray}
{\cal M}=\tilde{\cal M}
=
\frac{h}{1+e^{2 \Lambda x}}, 
\quad 
{\cal D}=\tilde{\cal D}
=
\frac{h}{1+e^{- 2 \Lambda x}}, 
\quad 
T
=
- \Lambda \tanh \Lambda x.
\label{Kap}
\end{eqnarray}
They also studied the domain wall for general values of the coupling 
$h^2 \not= 2\Lambda^2$ 
numerically and found that the qualitative features are unchanged. 

If we add a single flavor of quarks in the fundamental representation 
in the  ${\cal N}=2$ $SU(2)$ gauge theory, 
we obtain three singularities in the moduli space. 
{}For large bare mass of the quark, the additional singularity corresponds to 
the situation where the effective mass of quark vanishes, whereas  
the $Z_3$ symmetry among three singularities is realized in the limit of 
vanishing bare quark mass \cite{SeibergWitten}. 
These three singularities become three discrete vacua of ${\cal N}=1$ gauge 
theory when perturbed by the adjoint scalar mass \cite{IntriligatorSeiberg2}. 
In view of these features, we extend the $U(1)\times U(1)'$ model 
of ref.\cite{KSY} 
by adding an additional pair of chiral superfields ${\cal Q}, \tilde{\cal Q}$ 
corresponding to the quark and anti-quark 
\begin{eqnarray}
\begin{array}{ccc}
      & {\cal Q} & \tilde{\cal Q} \\
U(1)  &  1       &  -1            \\
U(1)' &  0       &  0   
\end{array}
\end{eqnarray} 
To make the quark massless at $T=m$ where $m$ is the bare mass 
parameter 
for the quark ${\cal Q}$, the superpotential is extended as 
\begin{equation}
W=(T-\Lambda) {\cal M} \tilde{{\cal M}}
+ (T+\Lambda) {\cal D} \tilde{{\cal D}}
+ (T-m) {\cal Q} \tilde{{\cal Q}}
- h^2 T. 
\end{equation}
This simple modification produces a model which 
possesses three distinct ${\cal N}=1$ supersymmetric vacua and allows us to 
obtain an exact solution for junctions. 
Since the action is invariant under the three global $U(1)$ transformations 
\begin{equation}
{\cal M} \to e^{i \delta_{1}} {\cal M}, \quad
\tilde{\cal M} \to e^{-i \delta_{1}} \tilde{\cal M}, \quad
{\cal D} \to e^{i \delta_{2}} {\cal D}, \quad
\tilde{\cal D} \to e^{-i \delta_{2}} \tilde{\cal D}, \quad
{\cal Q} \to e^{i \delta_{3}} {\cal Q}, \quad
\tilde{\cal Q} \to e^{-i \delta_{3}} \tilde{\cal Q}, \quad
\end{equation}
we can choose the vacuum configuration to be 
\begin{eqnarray}
{\rm Vac.1} &\!\!\!:&\!\!\!
T=m, \quad {\cal Q}=\tilde{\cal Q}= h, \quad
{\cal M}=\tilde{\cal M}={\cal D}=\tilde{\cal D}=0, \nonumber \\
{\rm Vac.2} &\!\!\!:&\!\!\!
T=\Lambda, \quad {\cal M}=\tilde{\cal M} = h, \quad
{\cal Q}=\tilde{\cal Q}={\cal D}=\tilde{\cal D}=0, \nonumber \\
{\rm Vac.3} &\!\!\!:&\!\!\!
T=-\Lambda, \quad {\cal D}=\tilde{\cal D}= h, \quad
{\cal Q}=\tilde{\cal Q}={\cal M}=\tilde{\cal M}=0.
\label{vac}
\end{eqnarray}

We will consider a field configuration which is static and translationally 
invariant along $x_3$ direction. 
We assume that the three different vacua are realized in different 
directions at spatial infinity in $x_1, x_2$ plane. 

\vspace{5mm}

{\large {\bf The solution }}

\vspace{5mm}

The states which are saturated by the 
Bogomol'nyi bound obey the eqs. (\ref{Be1}) and (\ref{Be1vector}) 
or the eqs. (\ref{Be2}) and (\ref{Be2vector}).
{}For simplicity, we now choose the eqs. (\ref{Be2}) and  (\ref{Be2vector}). 
The BPS equation (\ref{Be2vector}) for $U(1)\times U(1)'$ vector superfields 
can be satisfied 
trivially\footnote{
The $Z_3$ symmetric case of the vanishing bare quark mass in the 
Seiberg-Witten theory yields a 
different charge assignment for the third singularity $(n_m, n_e) =(1, 2)$ 
instead of $(n_m, n_e) =(0, 1)$ \cite{SeibergWitten}. 
Even if we use this charge assignment for the $Q$ field, 
The vanishing $D$ term condition gives the same result. 
 } 
 by $v_{\mu}=0$ and $D=0$. The other BPS equation (\ref{Be2}) becomes
\begin{equation}
2 K_{ij^*} \frac{ \partial A^i }{ \partial z}
=  \Omega \frac{\partial W^*}{\partial A^{*j}},
\quad 
\Omega = -i \frac{\langle i Z_1^* + Z_2^* \rangle}%
{|\langle i Z_1^* + Z_2^* \rangle |}.
\label{BPSeqs}
\end{equation}
We will look for a solution of this partial differential equation. 
We observe that the phase of $\Omega$ can be 
 absorbed by a 
rotation of field configuration, 
since the BPS equation (\ref{BPSeqs}) is invariant under a 
phase rotation
: 
$\Omega \to {\rm e}^{i \delta} \Omega, \ z \to {\rm e}^{-i \delta} z$. 
Later, we will check that the solution satisfies 
$H_{\rm II} > H_{\rm I}$.

Since we assume a canonical K\"ahler metric $K_{ij*}=\delta_{ij}$, 
the BPS eqs.~(\ref{BPSeqs}) become for our model 
\begin{eqnarray}
2 \frac{\partial {\cal M}}{\partial z} 
\!\!\!&=\!\!\!& 
\Omega \tilde{\cal M}^* 
(T-\Lambda)^*, \nonumber \\
2 \frac{\partial {\cal D}}{\partial z} 
\!\!\!&=\!\!\!&
\Omega \tilde{\cal D}^* 
(T+\Lambda)^*, \nonumber \\
2 \frac{\partial {\cal Q}}{\partial z} 
\!\!\!&=\!\!\!& 
\Omega \tilde{\cal Q}^* 
(T-m)^*, \nonumber \\
2 \frac{\partial T}{\partial z}
\!\!\!&=\!\!\!& 
\Omega
\left(
{\cal M}\tilde{\cal M} +
{\cal D}\tilde{\cal D} +
{\cal Q}\tilde{\cal Q} 
- h^2
\right)^*.
\label{4BPSeqs}
\end{eqnarray}

Eqs. (\ref{vac}) and (\ref{4BPSeqs}) are invariant under the following 
global phase changes of parameters, fields and complex coordinate $z$ 
\begin{eqnarray}
&
h \to e^{i \beta} h, \quad {\cal M} \to e^{i \beta} {\cal M}, \quad 
\tilde{\cal M} \to e^{i \beta} \tilde{\cal M}, 
& \nonumber \\
&
{\cal D} \to e^{i \beta} {\cal D}, \quad
\tilde{\cal D} \to e^{i \beta} \tilde{\cal D}, \quad
{\cal Q} \to e^{i \beta} {\cal Q}, \quad
\tilde{\cal Q} \to e^{i \beta} \tilde{\cal Q}, 
& \nonumber \\
&
\Lambda \to e^{i \gamma} \Lambda, 
\quad 
T \to e^{i \gamma} T,\quad 
z \to e^{2 i \beta + i \gamma} z. &
\end{eqnarray}
Arbitrary complex parameters $h$ and $\Lambda$ can be obtained 
from real-positive $h$ and $\Lambda$ by these phase changes. 
Therefore we shall take $h$ and $\Lambda$ to be real-positive in the 
following without loss of generality.  
The BPS condition $D=0$ yields 
$|{\cal M}(z, \bar z)|=|\tilde{\cal M}(z, \bar z)|$, 
$|{\cal D}(z, \bar z)|=|\tilde{\cal D}(z, \bar z)|$, and 
$|{\cal Q}(z, \bar z)|=|\tilde{\cal Q}(z, \bar z)|$.
Inspired by this condition, we wish to find a solution assuming 
\begin{equation}
{\cal M}(z, \bar z)=\tilde{\cal M}(z, \bar z), \quad
{\cal D}(z, \bar z)=\tilde{\cal D}(z, \bar z), \quad
{\cal Q}(z, \bar z)=\tilde{\cal Q}(z, \bar z).
\end{equation}
and that all of them are real-positive in the entire complex plane. 
We shall see that this Ansatz gives a consistent solution. 


We note that the model acquires a $Z_3$ symmetry if we choose 
the bare mass $m$ of ${\cal Q}$ as 
\begin{equation}
m = i \sqrt{3} \Lambda. 
\end{equation}
In order to obtain the exact analytic solution 
of the domain wall junction, 
we specialize to this case, 
and shift the field $T$ as 
$T^\prime = T - i \frac{1}{\sqrt{3}} \Lambda
$ 
 to make $T'=0$ as the origin of the $Z_3$ rotation 
$T' \to {\rm e}^{\pm i{2\pi \over 3}}T'$. 
The three vacua (\ref{vac}) and BPS equations (\ref{4BPSeqs}) 
take manifestly $Z_3$ symmetric forms 
\begin{eqnarray}
{\rm Vac.1} &\!\!\!:&\!\!\!
T^\prime=\frac{2}{\sqrt{3}} e^{i \frac{1}{2} \pi} \Lambda, \quad 
{\cal Q}=\tilde{\cal Q}=h, \quad
{\cal M}=\tilde{\cal M}={\cal D}=\tilde{\cal D}=0, \nonumber \\
{\rm Vac.2} &\!\!\!:&\!\!\!
T^\prime=\frac{2}{\sqrt{3}} e^{-i \frac{1}{6} \pi} \Lambda, \quad 
{\cal M}=\tilde{\cal M}=h, \quad
{\cal Q}=\tilde{\cal Q}={\cal D}=\tilde{\cal D}=0, \nonumber \\
{\rm Vac.3} &\!\!\!:&\!\!\!
T^\prime=\frac{2}{\sqrt{3}} e^{-i \frac{5}{6} \pi} \Lambda, \quad 
{\cal D}=\tilde{\cal D}=h, \quad
{\cal Q}=\tilde{\cal Q}={\cal M}=\tilde{\cal M}=0, 
\end{eqnarray}
\begin{eqnarray}
2 \frac{\partial}{\partial z} \ln q_M
&\!\!\!=&\!\!\!
\Omega 
\left(
{T^\prime}^* - \frac{2}{\sqrt{3}} e^{i \frac{1}{6} \pi} \Lambda
\right), \nonumber \\
2 \frac{\partial}{\partial z} \ln q_D
&\!\!\!=&\!\!\!
\Omega 
\left(
{T^\prime}^* - \frac{2}{\sqrt{3}} e^{i \frac{5}{6} \pi} \Lambda
\right), \nonumber \\
2 \frac{\partial}{\partial z} \ln q
&\!\!\!=&\!\!\!
\Omega 
\left(
{T^\prime}^* - \frac{2}{\sqrt{3}} e^{-i \frac{1}{2} \pi} \Lambda
\right), \nonumber \\
2 \frac{\partial}{\partial z} T^\prime
&\!\!\!=&\!\!\!
\Omega \,
h^2
\left(
q_M^2 + q_D^2 + q^2 -1
\right), 
\label{4BPSeqsV2}
\end{eqnarray}
where we have normalized the scalar fields by the nonzero expectation 
value  $h$ at vacua 
\begin{equation}
{\cal M}(z, \bar{z})=h \, q_M(z, \bar{z}), \quad 
{\cal D}(z, \bar{z})=h \, q_D(z, \bar{z}), \quad 
{\cal Q}(z, \bar{z})=h \, q(z, \bar{z}).
\end{equation}
The first of eq.(\ref{4BPSeqsV2}) can be rewritten as 
\begin{equation}
q_M = C(\bar{z}) \exp 
\left(
\frac{1}{2} \eta 
- \frac{1}{\sqrt{3}} \Omega e^{i \frac{1}{6} \pi } \Lambda z
\right),
\end{equation}
\begin{equation}
\frac{\partial}{\partial z} \eta({z, \bar{z}})
= \Omega \, {T^\prime}^*({z, \bar{z}}),
\label{eta}
\end{equation}
where the unknown function $C(\bar z)$ is determined by the reality 
condition 
for $q_M$ up to a constant which is absorbed into $\eta$ 
\begin{equation}
C(\bar{z})= \exp 
\left(
- \frac{1}{\sqrt{3}} \Omega^* e^{-i \frac{1}{6}\pi} \Lambda \bar{z}
\right). 
\end{equation}
The remaining unknown function $\eta(z, \bar{z})$ should then be real. 
Consequently we obtain 
\begin{equation}
q_M = \exp
\left(
\frac{1}{2} \eta 
+ \frac{2}{\sqrt{3}} \Lambda
{\rm Re} \left( -\Omega \, e^{i \frac{1}{6} \pi} z \right) 
\right), \quad 
\eta(z, \bar{z})= \left(\eta(z, \bar{z})\right)^*. 
\end{equation}
By an exactly similar procedure, we solve the second and third equations and 
obtain 
\begin{eqnarray}
q_D 
&\!\!\!=&\!\!\! 
\exp
\left(
\frac{1}{2} \eta 
+ \frac{2}{\sqrt{3}} \Lambda
{\rm Re} \left( - \Omega \, e^{i \frac{5}{6} \pi} z \right) 
+C_D \right), 
\qquad C_D \in {\mathbb R}, \\
q
&\!\!\!=&\!\!\! 
\exp
\left(
\frac{1}{2} \eta 
+ \frac{2}{\sqrt{3}} \Lambda
{\rm Re} \left( - \Omega \, e^{-i \frac{1}{2} \pi} z \right) 
+C \right),
\qquad C \in {\mathbb R},
\end{eqnarray}
where $C_D$ and $C$ are integration constants. 
Let us assume that the origin $z=0$ is the center of the domain wall junction 
and is $Z_3$ symmetric. 
Therefore, $q_M=q_D=q$ at $z=0$, which implies 
$C_D = C = 0$. 
Inserting eq.(\ref{eta}) to the complex conjugate of the last of 
eq.(\ref{4BPSeqsV2}), 
we obtain 
\begin{eqnarray}
2 \frac{\partial^2 }{\partial z \partial \bar{z}} \eta
&\!\!\!=&\!\!\!
- h^2 
\Biggl[
1 - e^{\eta 
}
\Biggl\{
\exp \left(
\frac{4 \Lambda}{\sqrt{3}}
{\rm Re} \left( - \Omega \, e^{i \frac{1}{6} \pi} z \right) \right)
\nonumber \\
&\!\!\! &\!\!\!
{}+
\exp \left(
\frac{4 \Lambda}{\sqrt{3}}
{\rm Re} \left( - \Omega \, e^{i \frac{5}{6} \pi} z \right) \right)
+\exp \left(
\frac{4 \Lambda}{\sqrt{3}}
{\rm Re} \left( - \Omega \, e^{-i \frac{1}{2} \pi} z \right) \right)
\Biggr\}
\Biggr].
\label{nijisiki}
\end{eqnarray}

{}For the special case of $h^2=2 \Lambda^2$, 
 eq. (\ref{nijisiki}) can be solved analytically. 
Imposing the boundary conditions at infinity we obtain the solution 
\begin{eqnarray}
\eta(z,\bar{z})
&\!\!\!=&\!\!\!
-2 \ln
\Biggl[
\exp \left(
\frac{2 \Lambda}{\sqrt{3}}
{\rm Re} \left( - \Omega \, e^{i \frac{1}{6} \pi} z \right) \right)
\nonumber \\ &\!\!\! &\!\!\! {} 
+\exp \left(
\frac{2 \Lambda}{\sqrt{3}}
{\rm Re} \left( - \Omega \, e^{i \frac{5}{6} \pi} z \right) \right)
+\exp \left(
\frac{2 \Lambda}{\sqrt{3}}
{\rm Re} \left( - \Omega \, e^{-i \frac{1}{2} \pi} z \right) \right)
\Biggr].
\end{eqnarray}
Therefore we find solutions for scalar fields as 
\begin{eqnarray}
{\cal M}(z, \bar{z})
&\!\!\! = &\!\!\! 
\tilde{\cal M}(z, \bar{z})=
\frac{\sqrt2 \Lambda s}{s+t+u}, \nonumber \\
{\cal D}(z, \bar{z})
&\!\!\! = &\!\!\! 
\tilde{\cal D}(z, \bar{z})=
\frac{\sqrt2 \Lambda t}{s+t+u}, \nonumber \\
{\cal Q}(z, \bar{z})
&\!\!\! = &\!\!\! 
\tilde{\cal Q}(z, \bar{z})=
\frac{\sqrt2 \Lambda u}{s+t+u}, \nonumber \\
T^\prime(z, \bar{z})
&\!\!\! = &\!\!\! 
\frac{2 \Lambda}{\sqrt{3}}
\frac{
e^{-i \frac{1}{6} \pi} s
+ e^{-i \frac{5}{6} \pi} t
+ e^{i \frac{1}{2} \pi} u
}{s+t+u}, 
\label{exact solution}
\end{eqnarray}
\begin{eqnarray}
s
&\!\!\!=&\!\!\!
\exp \left(
\frac{2 \Lambda}{\sqrt{3}}
{\rm Re} \left( - \Omega \, e^{i \frac{1}{6} \pi} z \right) \right),
\nonumber \\
t
&\!\!\!=&\!\!\!
\exp \left(
\frac{2 \Lambda}{\sqrt{3}}
{\rm Re} \left( - \Omega \, e^{i \frac{5}{6} \pi} z \right) \right),
\nonumber \\
u
&\!\!\!=&\!\!\!
\exp \left(
\frac{2 \Lambda}{\sqrt{3}}
{\rm Re} \left( - \Omega \, e^{-i \frac{1}{2} \pi} z \right) \right).
\label{stu}
\end{eqnarray}

\begin{figure}[htbp]
\begin{center}
\includegraphics{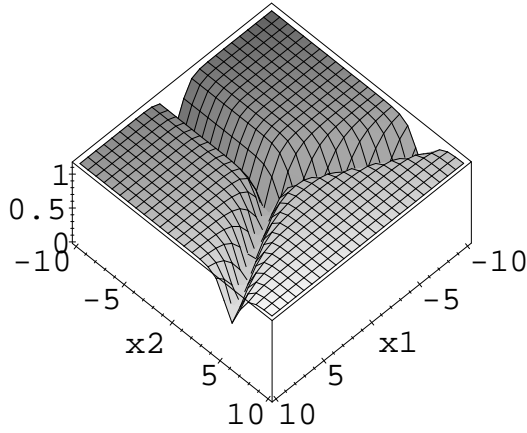}
\caption{The modulus of the field $T'$ as a function of $x_1$ and $x_2$. 
We set $\Lambda =1$ for simplicity. 
}
\label{FIG:1}
\end{center}
\end{figure}

Now we will examine the solution more closely. 
The domain wall separating vacua $I$ and $J$ is characterized by 
a normal vector directing from $I$ to $J$ which is 
expressed as a complex number of unit modulus $\omega_{I J}$. 
If the difference of the superpotential $W({\rm Vac}. I)$ at the vacuum $I$ 
and  $W({\rm Vac}. J)$ at $J$ is 
denoted as $\Delta W_{IJ}=W({\rm Vac}. J)-W({\rm Vac}. I)$, 
the integral form of the BPS equation gives the condition on the 
direction of the domain walls as \cite{CHT} 
\begin{equation}
\Omega \, \frac{\Delta {W_{IJ}}^*}{\left|\Delta W_{IJ}\right|}
\, \omega_{IJ}=1. 
\end{equation}
To orient the domain wall separating the vacuum 2 and 3 along the negative 
$x_2$ axis, we choose $\Omega=-1$. 
The modulus of the field $T'$ is plotted as a function of $x_1$ and $x_2$ in 
 {}Fig. \ref{FIG:1} 
where we can recognize three valleys corresponding to three domain walls. 

Let us first examine the boundary conditions at spatial infinity 
 $|z| \to \infty $.
{}From eqs. (\ref{exact solution}) and (\ref{stu}), we find 
\begin{equation}
\begin{array}{crcll}
{\rm when} & -\frac{1}{2}\pi < \arg(z) < \frac{1}{6}\pi, &
{\rm then}& s \gg t, u, & 
T^\prime \to \frac{2}{\sqrt{3}} e^{-i \frac{1}{6} \pi} \Lambda, \quad 
({\rm vac.2})
\\
{\rm when} & \frac{1}{6}\pi < \arg(z) < \frac{5}{6}\pi, &
{\rm then}& u \gg s, t, & 
T^\prime  \to \frac{2}{\sqrt{3}} e^{i \frac{1}{2} \pi} \Lambda, \quad 
({\rm vac.1})
\\
{\rm when} & \frac{5}{6}\pi < \arg(z) < \frac{3}{2}\pi, &
{\rm then}& t \gg s, u, & 
T^\prime \to \frac{2}{\sqrt{3}} e^{-i \frac{5}{6} \pi} \Lambda, \quad 
({\rm vac.3})
\end{array} 
\label{BC}
\end{equation}
Secondly, 
let us examine the asymptotic behavior along the region between two 
neighboring vacua. 
In the limit $x_2 \to - \infty$ with fixed $x_1$, eq.(\ref{exact solution}) 
reduces to 
\begin{eqnarray}
{\cal M} 
&\!\!\! \to &\!\!\!
\frac{\sqrt2 \Lambda e^{\Lambda x_1}}{e^{\Lambda x_1}+e^{-\Lambda x_1}}, 
\quad 
{\cal D} 
\to 
\frac{\sqrt2 \Lambda e^{-\Lambda x_1}}{e^{\Lambda x_1}+e^{-\Lambda x_1}},
\quad 
{\cal Q} 
\to 
0, 
\nonumber \\
T^\prime 
&\!\!\! \to &\!\!\!
\Lambda \tanh \Lambda x_1 
- \frac{\Lambda}{\sqrt{3}} i.
\label{ylimit}
\end{eqnarray}
Thus we recover the exact solution of domain wall (\ref{Kap}) 
with $x$ replaced by $-x_1$. 
By the $Z_3$ symmetry, we also obtain respective exact domain wall 
solutions at the asymptotic region $x_1 = \pm \sqrt3 x_2$ correctly. 

{}Finally let us evaluate the central charges 
$\langle Z_k \rangle$ and $\langle Y_k \rangle$ and 
check $H_{\rm II} > H_{\rm I}$ to confirm that this solution is indeed 
realized as a $1/4$ BPS state. 
Since these charges are determined solely by the boundary condition 
at spatial infinity (\ref{BC}), we evaluate them on a large 
cylindrical region with a disk of large radius $R$ ($ R \gg \Lambda^{-1}$) 
centered at $z=0$ 
and a height $\Delta x_3$. 
Field configurations on the surface 
of the large cylinder approaches 
a step-function across domain walls. 
We find 
\begin{eqnarray}
\langle Z_1 \rangle
 = 
-12 \Lambda^3 R \Delta x_3, 
\quad 
\langle Z_2 \rangle
 = 
i 12 \Lambda^3 R \Delta x_3, 
\quad 
\langle Y_3 \rangle
 = 
-2 \sqrt{3} \Lambda^2 \Delta x_3,
\label{ex.val.}
\end{eqnarray}
with corrections suppressed exponentially as $R \to \infty$. 
Therefore we obtain 
\begin{eqnarray}
H_{\rm I} 
&\!\!\! = &\!\!\! 
|\langle -i Z_1-Z_2 \rangle|
-\langle Y_3 \rangle 
= 2\sqrt3\Lambda^2 \Delta x_3, 
\nonumber \\
H_{\rm II} 
&\!\!\! = &\!\!\! 
|\langle i Z_1-Z_2 \rangle|
+\langle Y_3 \rangle
=
24 \Lambda^3 R \Delta x_3 
- 2\sqrt3\Lambda^2 \Delta x_3.
\label{massofjunction}
\end{eqnarray}
We see that $H_{\rm II} > H_{\rm I}$ confirming the correctness of the 
choice of the BPS equation (\ref{Be2}). 
It is interesting to observe that $H_{\rm II}$ is larger than $H_{\rm I}$ 
primarily due to the different phases of $\langle Z_1\rangle$ and 
$\langle Z_2 \rangle$ and 
not to the presence of $\langle Y_3 \rangle \not=0$ term. 
This is in contrast to the case of a single domain wall where 
$H_{\rm I}=H_{\rm II}$ since $\langle Z_1\rangle$ and 
$\langle Z_2 \rangle$ have the same phase factor and 
$\langle Y_3 \rangle =0$. 
In fact we observe in eq.(\ref{massofjunction}) 
that the contribution of $\langle Y_3 \rangle $ to the 
mass of the domain wall junction is actually negative\footnote{
This fact seems to be against previous thoughts such as in 
ref.\cite{GibbonsTownsend}. }. 
To see this fact from another viewpoint, 
let us consider the central charge $\langle Y_3 \rangle$ further. 
The general formula (\ref{centralchargeY}) 
for the case of many chiral superfields can be partially integrated as 
\begin{equation}
Y_3 =
\int d x_3 \,
i  
\int d^2 x \,
\left[\partial_1\left(K_{i}
 \partial_2 A^i\right)
-\partial_2\left(K_{i}
 \partial_1 A^i\right)
\right] 
=
\int d x_3 \,
i  
\oint K_{i} d A^i, 
\label{y-area-generic}
\end{equation}
where $K_i \equiv \partial K / \partial A^i$ and the last integral 
 in the field space should be done as a map from a counter clockwise 
 contour in the $z$ plane. 
In our case, we have contributions to $\langle Y_3 \rangle$ from 
the field $T$ only, since eq.(\ref{centralchargeY}) clearly shows 
that fields with real values do not contribute. 
Moreover the K\"ahler metric in our case is trivial and the counter clockwise 
contour in $z$ is mapped to a counter clockwise contour in the field 
$T$. 
Therefore we obtain 
\begin{equation}
Y_3 =
\int d x_3 \,
(-2) 
\int d({\rm Re}T) d({\rm Im}T) 
=
\int d x_3 \,
(-2) \sqrt3 \Lambda^2, 
\end{equation}
where the integration region in the field space $T$ is the 
equilateral triangle whose vertices are the three vacuum field values. 
We see that the central charge $\langle Y_3 \rangle$ has 
a simple geometrical meaning of the $-2$ times the triangular 
area in field space which is 
enclosed by three domain walls connecting three distinct vacua at infinity. 
From this consideration, we again find that the central charge 
 $\langle Y_3 \rangle$ should be negative and does not have a 
naive meaning of ``junction mass''. 
Let us also note that the domain walls correspond to straight lines 
in field space in our simple model.  
For general models, it has been shown that lines corresponding to 
domain walls are not straight lines in field space $A^i$, but become 
straight lines if mapped to the complex plane of superpotential $W(A^i)$ 
\cite{CHT}, \cite{Saffin}. 
Therefore the geometrical meaning of the central charge 
 $\langle Y_3 \rangle$ in general situation (\ref{y-area-generic}) 
is that it is proportional to 
the area in field space spanned by the fields as measured 
by the K\"ahler potential \cite{CHT}. 

Let us emphasize that the central charge $\langle Y_3 \rangle$ 
 has a simple geometrical meaning and is negative in our model. 
The main contribution to the mass of the domain wall junction configuration 
comes from the central charge $Z_k$ and the negative $\langle Y_3 \rangle$ 
is merely an additional small negative contribution. 
This result is not an artifact of our choice of the BPS 
equation (\ref{Be2}) rather than the other possibility (\ref{Be1}). 
If we choose the other BPS equation, we merely obtain the reflected 
domain wall junction solution $x_1 \to -x_1, x_2 \to x_2$. 
The solution in fact gives a positive $\langle Y_3 \rangle$, but it also 
accompanies a different formula for the mass of the configuration 
$\langle H \rangle=H_{\rm I} \equiv
|\langle -i Z_1-Z_2 \rangle|
-\langle Y_3 \rangle$ in 
(\ref{BPSenerges}) where 
the central charge $\langle Y_3 \rangle$ contributes negatively to 
the mass. 
Therefore the final physical result is identical. 

When we complete writing our paper, more new works appeared on domain 
walls and junctions \cite{GabadadzeShifman}.


One of the authors (H.O.) gratefully acknowledges support from 
the Iwanami Fujukai Foundation.
This work is supported in part by Grant-in-Aid 
for Scientific Research from the Japan Ministry 
of Education, Science and Culture for 
the Priority Area 291 and 707.


\end{document}